\documentclass[noendmarks,draft]{arxiv}

\pdfoutput=1

\newif\iffull
\fulltrue

\usetikzlibrary{arrows,automata}
\tikzset{automaton/.style={
	auto,shorten >=1pt,node distance=2cm,>=stealth',
	every state/.style={minimum size=2.6em,inner sep=0pt},
    every node/.style={font=\scriptsize},
	initial/.style={initial by arrow,initial text=,initial distance=0.5cm}
}}

\newcommand{\Cerny}{\v{C}ern\'{y}\xspace}
\newcommand{\A}{\ensuremath{\mathcal{A}}}

\newcommand{\low}{\mathit{low}}
\newcommand{\high}{\mathit{high}}

\newcommand{\ShortRW}{\mbox{\textsc{short-reset-word}}\xspace}
\newcommand{\ShortestRW}{\mbox{\textsc{shortest-reset-word}}\xspace}
\newcommand{\SAT}{\mbox{\textsc{sat}}\xspace}
\newcommand{\TwoSAT}{\mbox{\oldstylenums{2}\textsc{sat}}\xspace}
\newcommand{\SharpSAT}{\mbox{\#\textsc{sat}}\xspace}

\newcommand{\SATUNSAT}{\mbox{\textsc{sat-unsat}}\xspace}
\newcommand{\FSAT}{\mbox{\textsc{fsat}}\xspace}
\newcommand{\MAXSAT}{\mbox{\textsc{max-sat}}\xspace}
\newcommand{\MAXSATSIZE}{\mbox{\textsc{max-sat-size}}\xspace}
\newcommand{\MAXWEIGHTSAT}{\mbox{\textsc{max-weight-sat}}\xspace}

\newcommand{\true}{\mbox{true}}
\newcommand{\false}{\mbox{false}}

\newcommand{\LOGSpace}{\textsc{Logspace}\xspace}
\newcommand{\FP}{\textup{FP}\xspace}
\newcommand{\SharpP}{\textup{{\#}P}\xspace}
\newcommand{\NPMV}{\textup{NPMV}\xspace}
\newcommand{\NPMVg}{\ensuremath{\text{NPMV}\mkern-3mu_g}\xspace}
\newcommand{\FPNP}{\ensuremath{\FP^\NP}\xspace}
\newcommand{\FPNPlog}{\ensuremath{\FP^{\NP[\log]}}\xspace}

\newcommand{\M}[2]{\ensuremath{\Sigma^{#2}_{#1}}}
\newcommand{\N}[2]{\ensuremath{\Gamma^{#2}_{#1}}}
\newcommand{\Q}[2]{\ensuremath{q_{#1,#2}}}
\renewcommand{\P}[2]{\ensuremath{p_{#1,#2}}}
\newcommand{\R}[2]{\ensuremath{r_{#1,#2}}}

\let\qednow\qed

\title{The Complexity of Finding Reset Words\\ in Finite Automata}
\author{J\"org Olschewski\addr{1}\fnmsep\addr{2}\fnmsep\thanks{supported by the ESF project GASICS.} and Michael Ummels\addr{2}\fnmsep\addr{3}\fnmsep\thanks{supported by the French project DOTS (ANR-06-SETI-003).}}

\address{Lehrstuhl Informatik~7, RWTH Aachen University, Germany \\
\email{olschewski@automata.rwth-aachen.de}}

\address{LSV, CNRS \& ENS Cachan, France}

\address{Mathematische Grundlagen der Informatik, \\ 
RWTH Aachen University, Germany \\
\email{ummels@logic.rwth-aachen.de}}

\begin{document}

\maketitle

\begin{abstract}
We study several problems related to finding reset words in deterministic finite
automata. In particular, we establish that the problem of deciding whether a shortest
reset word has length~$k$ is complete for the complexity class \DP. This result
answers a question posed by Volkov.
For the search problems of finding a shortest reset word and the length of a
shortest reset word, we establish membership in the complexity classes \FPNP and
\FPNPlog, respectively. Moreover, we show that both these problems are hard for
\FPNPlog. Finally, we observe that computing a reset word of a given length is
\FNP-complete.
\end{abstract}

\section{Introduction}

A \emph{synchronising automaton} is a deterministic finite automaton that can
be reset to a single state by reading a suitable word. More precisely, we require
needs to exist a word~$w$ such that, no matter at which state of the automaton we
start, $w$~takes the automaton to the same state~$q$; we call any such word~$w$
a \emph{reset word} or a \emph{synchronising word}.
Although it is easy to decide whether a given automaton is synchronising and to
compute a reset word, finding a \emph{shortest} reset word seems to be
a hard problem.

The motivation to study reset words does not only come from automata theory: 
There are applications in the fields of many-valued logics,
biocomputing, set theory, and many more \cite{Volkov08}. A purely
mathematical viewpoint can be obtained by identifying letters with their
associated transition functions, which act on a finite set. The task is then to
find a composition of these functions such that the resulting
function is constant.

The theory of synchronising automata has been established in the 1960s and is
still actively developed. The famous \emph{\Cerny Conjecture} was formulated
in 1971 \cite{CPR71}. The conjecture claims that every synchronising automaton
with $n$~states has a reset word of length $(n-1)^2$. As of now, the conjecture
has neither been proved nor disproved; the best known upper bound on
the length of a reset word is $(n^3-n)/6$, as shown by Pin \cite{Pin83}.

While Eppstein~\cite{Eppstein90} showed that the problem of deciding whether
there exists a reset word of a given length~$k$ is \NP-complete, the complexity
of deciding whether a \emph{shortest} reset word has length~$k$ is not known
to be in \NP. In his survey paper \cite{Volkov08}, Volkov asked for
the precise complexity of this problem.
In this paper, we show that deciding whether a shortest reset word has
length~$k$ is complete for the class \DP, the closure of $\NP\cup\coNP$
under finite intersections.
In particular, since every \DP-complete problem is both
\NP-hard and \coNP-hard, it is unlikely that the problem of deciding the
length of a shortest reset word lies in $\NP\cup\coNP$.\footnote{
We have been informed
that Gawrychowski \cite{Gawrychowski08}
has shown \DP-completeness of \ShortestRW earlier, but his proof has never
been published. While his reduction uses a five-letter alphabet, we prove
hardness even over a binary alphabet.}

The class \DP is contained in the class \PNP, i.e.\ every problem in \DP can be
solved by a deterministic polynomial-time Turing machine that has access to an
oracle for an \NP-complete problem. In fact, two oracle queries suffice
for this purpose. If one restricts the number of oracle queries to be
logarithmic in the size of the input, one arrives at the class \PNPlog, which
is believed to be a proper superclass of \DP. We show that the problem of
computing the length of a shortest reset word (as opposed to deciding whether
it is equal to a given integer) is, in fact, complete for \FPNPlog, the
functional analogue of \PNPlog. Hence, this problem seems to be even harder
than deciding the length of a shortest reset word. Our result complements a
recent result by Berlinkov \cite{Berlinkov10}, who showed that, unless
$\PTime=\NP$, there is no polynomial-time algorithm that \emph{approximates}
the length of a shortest reset word within a constant factor.

For the more general problem of computing a shortest reset word (not only its
length), we prove membership in \FPNP, the functional analogue of \PNP.
While our lower bound of \FPNPlog on computing the length of a shortest reset
word carries over to this problem, we leave it as an open problem whether
computing a shortest reset word is also \FPNP-hard.

Apart from studying problems related to computing a \emph{shortest} reset word,
we also consider the problem of computing a reset word of a given length
(represented in unary). We observe that this problem is complete for the class
\FNP of search problems for which a solution can be verified in polynomial time.
In other words: the problem is as hard as computing a satisfying assignment for
a given Boolean formula.

\section{Preliminaries}

Let $\A=\langle Q,\Sigma,\delta\rangle$ be a deterministic finite automaton
(DFA)
with finite state set~$Q$, finite alphabet~$\Sigma$ and transition function
$\delta\colon Q\times\Sigma\to Q$.
The transitive closure of~$\delta$ can be defined inductively by
$\delta^*(q,\epsilon)=q$ and $\delta^*(q,wa)=\delta(\delta^*(q,w),a)$
for each $q\in Q$, $w\in\Sigma^*$ and $a\in\Sigma$.
We call any word $w\in\Sigma^*$ such that $|\{\delta^*(q,w)\mid q\in Q\}|=1$
a \emph{reset word} for~$\A$, and we say that $\A$~is \emph{synchronising} if
such a word exists. Note that, if $w$~is a reset word for~$\A$, then so
is~$xwy$ for all $x,y\in\Sigma^*$.

\iffull
We assume that the reader is familiar with basic concepts of complexity theory,
in particular with the classes \PTime, \NP and \coNP. We will introduce the other
complexity classes that play a role in this paper on the fly;
see \cref{appendix:complexity} for formal definitions.
\else
We assume that the reader is familiar with basic concepts of complexity theory,
in particular with the classes \PTime, \NP and \coNP. We will introduce the other
classes that play a role in this paper on the fly. For details, see
\cite{Papadimitriou94,Selman94}.
\fi

\section{Decision Problems}

The most fundamental decision problem concerning reset words is
to decide whether a given deterministic finite automaton is synchronising.
\Cerny~\cite{Cerny64} noted that it suffices to check for each pair
$(q,q')$ of states whether there exists a word~$w\in\Sigma^*$ with
$\delta^*(q,w)=\delta^*(q',w)$. The latter property can obviously be decided
in polynomial time. The best known algorithm for \emph{computing} a
reset word is due to Eppstein~\cite{Eppstein90}: his
algorithm runs in time $\Oh(|Q|^3+|Q|^2\cdot |\Sigma|)$.
Computing a \emph{shortest} reset word, however, cannot be done in
polynomial time unless the following decision problems are in $\PTime$.

\begin{quote}
\ShortRW: Given a DFA~$\A$ and a positive integer~$k$,
decide whether there exists a reset word for~$\A$ of length~$k$.
\end{quote}
\begin{quote}
\ShortestRW: Given a DFA~$\A$ and a positive integer~$k$,
decide whether the minimum length of a reset word for~$\A$ equals~$k$.
\end{quote}

If the parameter~$k$ is given in unary, it is obvious that \ShortRW is in
\NP. However, even if $k$~is given in binary, this problem is in \NP:
since every synchronising automaton has a reset word of length
$p(|Q|)$ (where $p$ is a low-degree polynomial, e.g.\ $p(n)=(n^3-n)/6$),
to establish whether there exists a reset word of length~$k$, it suffices
to guess a reset word of length $\min\{p(|Q|),k\}$.
Eppstein~\cite{Eppstein90} gave a matching lower bound by proving that
\ShortRW is also \NP-hard.

Regarding \ShortestRW, Samotij~\cite{Samotij07} showed that the problem is
\NP-hard.
\iffull
He also claimed that \ShortestRW is \coNP-hard. However, to prove
\coNP-hardness, he reduced from the \emph{validity problem} for Boolean
formulae in 3CNF (i.e., the problem of deciding whether such a formula is a
tautology), which is decidable in polynomial time. (Note that a CNF formula
is valid if and only if each of its clauses contains both a positive and a
negative occurrence of a variable).
\fi
We prove that \ShortestRW is complete for \DP, the
class of all languages of the form $L=L_1\setminus L_2$ with $L_1,L_2\in\NP$.
Since \DP is a superclass of both \NP and \coNP, our result implies hardness
for both of these classes.
In fact, we show that \ShortestRW is \DP-hard even over a binary alphabet.

\begin{theorem}\label{thm:dp-complete}
\ShortestRW is \DP-complete.
\end{theorem}
\begin{proof}
It is easy to see that \ShortestRW belongs to \DP: indeed, we can
write \ShortestRW as the difference of
$\ShortRW$ and $\ShortRW^-$, where
\[\ShortRW^-=\{(\A,k+1)\mid (\A,k)\in\ShortRW\},\]
a problem which is obviously in \NP (even if $k$~is given in
binary).\pagebreak[1]

It remains to prove that \ShortestRW is \DP-hard. We reduce from the
canonical \DP-complete problem \SATUNSAT: given two Boolean formulae $\phi$
and~$\psi$ (in CNF), decide whether $\phi$~is satisfiable and $\psi$~is
unsatisfiable. More precisely, we show how to construct (in polynomial time)
from a pair $(\phi,\psi)$ of Boolean formulae in CNF over propositional
variables $X_1,\ldots,X_k$ a synchronising automaton~$\A$ over the alphabet
$\Sigma=\{0,1\}$ with the following properties:
\begin{enumerate}
  \item If $\phi$ and~$\psi$ are satisfiable, then there exists a reset word of
length~$k+2$.
  \item If $\phi$~is satisfiable and $\psi$ is unsatisfiable, then a shortest
reset word has length~$k+3$.
  \item If $\phi$~is unsatisfiable, then every reset word has
    length at least $k+4$.
\end{enumerate}
From 1.--3.\ we get that $\phi$ is satisfiable and $\psi$ is unsatisfiable if
and only if a shortest reset word has length~$k+3$.

Given formulae $\phi=C_1\wedge\ldots\wedge C_n$ and
$\psi=D_1\wedge\ldots\wedge D_n$ where, without loss of generality, $\phi$
and~$\psi$ have the same number~$n$ of clauses, and no propositional variable
occurs in both $\phi$ and~$\psi$, the automaton~$\A$ consists of the states
$s$, $t_1$, $t_2$, $\P{i}{j}$ and $\Q{i}{j}$, $i\in\{1,\ldots,n\}$,
$j\in\{\bot,\top,1,\ldots,k\}$; the transitions
are depicted in \cref{fig:reduction-2}:
an edge from $p$ to~$q$ labelled with $\Sigma'\subseteq\Sigma$ has the
meaning that $\delta(p,a)=q$ for each $a\in\Sigma'$.
The sets $\M{i}{j}\subseteq\Sigma$
are defined by $0\in\M{i}{j}\Leftrightarrow\neg X_j\in C_i$ and
$1\in\M{i}{j}\Leftrightarrow X_j\in C_i$, and the sets
$\N{i}{j}\subseteq\Sigma$ are defined by $0\in\N{i}{j}\Leftrightarrow
\neg X_j\in D_i$ and $1\in\N{i}{j}\Leftrightarrow X_j\in D_i$.
Hence, e.g.\, $0\in\M{i}{j}$ if we can satisfy the $i$th clause of~$\phi$
by setting variable~$X_j$ to $\false$.

\begin{figure}[t]
\centering
\begin{tikzpicture}[automaton,x=0.9cm]

\node[state]	(C1bot) at (-2,0)	{\P{1}{\bot}};
\node[state]	(C1top) at (0,0)	{\P{1}{\top}};
\node[state]	(C1X1) at (2,0)		{\P{1}{1}};
\node			(C1X2) at (3.75,0)		{};
\node			(C1Xk-1) at (4.75,0)	{};
\node at (barycentric cs:C1X2=0.5,C1Xk-1=0.5) {$\cdots$};
\node[state]	(C1Xk) at (6.5,0)		{\P{1}{k}};

\node at (1,-2) {$\vdots$};

\node[state]	(Cmbot) at (-2,-4)	{\P{n}{\bot}};
\node[state]	(Cmtop) at (0,-4)	{\P{n}{\top}};
\node[state]	(CmX1) at (2,-4)		{\P{n}{1}};
\node			(CmX2) at (3.75,-4)		{};
\node			(CmXk-1) at (4.75,-4)	{};
\node at (barycentric cs:CmX2=0.5,CmXk-1=0.5) {$\cdots$};
\node[state]	(CmXk) at (6.5,-4)		{\P{n}{k}};

\node			(s1) at (8,-2)	{};
\node[state]	(t1) at (9.5,-2)	{$t_1$};


\begin{scope}[yshift=-7cm]

\node[state]	(D1bot) at (-2,0)	{\Q{1}{\bot}};
\node[state]	(D1top) at (0,0)	{\Q{1}{\top}};
\node[state]	(D1X1) at (2,0)		{\Q{1}{1}};
\node			(D1X2) at (3.75,0)		{};
\node			(D1Xk-1) at (4.75,0)	{};
\node at (barycentric cs:D1X2=0.5,D1Xk-1=0.5) {$\cdots$};
\node[state]	(D1Xk) at (6.5,0)		{\Q{1}{k}};

\node[state]	(Dmbot) at (-2,-4)	{\Q{n}{\bot}};
\node[state]	(Dmtop) at (0,-4)	{\Q{n}{\top}};
\node[state]	(DmX1) at (2,-4)		{\Q{n}{1}};
\node			(DmX2) at (3.75,-4)		{};
\node			(DmXk-1) at (4.75,-4)	{};
\node at (barycentric cs:DmX2=0.5,DmXk-1=0.5) {$\cdots$};
\node[state]	(DmXk) at (6.5,-4)		{\Q{n}{k}};

\node at (1,-2) {$\vdots$};

\node			(s2) at (8,-2)	{};
\node[state]	(t2) at (9.5,-2)	{$t_2$};
\end{scope}


\node			(middle) at (barycentric cs:t1=0.5,t2=0.5)		{};
\node[state]	(s) at (8.5,0 |- middle)		{$s$};


\path[->]	(C1bot)		edge[out=-50,in=-130,loop] node {$1$}		(C1bot)
			(C1bot)		edge	node {$0$}							(C1top)
			(C1top)		edge[out=-50,in=-130,loop] 	node {$0$}		(C1top)
			(C1top)		edge	node {$1$}							(C1X1)
			(C1X1)		edge	node[pos=0.6] {$\Sigma\setminus\M{1}{1}$}	(C1X2)
			(C1Xk-1)	edge	node[pos=0.4] {$\Sigma\setminus\M{1}{k-1}$}	(C1Xk)
			(C1Xk)		edge[bend left,pos=0.0625]	node {$\Sigma\setminus\M{1}{k}$}	(t1);
\path[->]	(Cmbot)		edge[out=50,in=130,loop] node[swap] {$1$}					(Cmbot)
			(Cmbot)		edge	node[swap] {$0$}							(Cmtop)
			(Cmtop)		edge[out=50,in=130,loop] node[swap] {$0$}					(Cmtop)
			(Cmtop)		edge	node[swap] {$1$}							(CmX1)
			(CmX1)		edge	node[swap,pos=0.6] {$\Sigma\setminus\M{n}{1}$}	(CmX2)
			(CmXk-1)	edge	node[swap,pos=0.4] {$\Sigma\setminus\M{n}{k-1}$}	(CmXk)
			(CmXk)		edge[bend right,swap,pos=0.0625]	node {$\Sigma\setminus\M{n}{k}$}	(t1);

\path[->]	(t1)			edge	node {$\Sigma$}		(t2);
\path[->]	(s)			edge[out=140,in=-140,loop,swap]	node {$\Sigma$}		(s);

\begin{scope}[out=-90,in=180,looseness=0.4,very near start,swap]
\path	(C1X1)		edge	node {$\M{1}{1}$}	(s1.center)
			(C1Xk)		edge[looseness=1]	node {$\M{1}{k}$}	(s1.center);
\end{scope}
\begin{scope}[out=90,in=180,looseness=0.4,very near start]
\path	(CmX1)		edge	node {$\M{n}{1}$}	(s1.center)
			(CmXk)		edge[looseness=1]	node {$\M{n}{k}$}	(s1.center);
\end{scope}

\path[->]	(D1bot)		edge[out=-50,in=-130,loop] node {$1$}		(D1bot)
			(D1bot)		edge	node {$0$}							(D1top)
			(D1top)		edge[out=-50,in=-130,loop]	node {$0$}		(D1top)
			(D1top)		edge	node {$1$}							(D1X1)
			(D1X1)		edge	node[pos=0.6] {$\Sigma\setminus\N{1}{1}$}	(D1X2)
			(D1Xk-1)	edge	node[pos=0.4] {$\Sigma\setminus\N{1}{k-1}$}	(D1Xk)
			(D1Xk)		edge[bend left,pos=0.0625]	node {$\Sigma\setminus\N{1}{k}$}	(t2);
\path[->]	(Dmbot)		edge[out=50,in=130,loop] node[swap] {$1$}					(Dmbot)
			(Dmbot)		edge	node[swap] {$0$}							(Dmtop)
			(Dmtop)		edge[out=50,in=130,loop]	node[swap] {$0$}				(Dmtop)
			(Dmtop)		edge	node[swap] {$1$}							(DmX1)
			(DmX1)		edge	node[swap,pos=0.6] {$\Sigma\setminus\N{n}{1}$}	(DmX2)
			(DmXk-1)	edge	node[swap,pos=0.4] {$\Sigma\setminus\N{n}{k-1}$}	(DmXk)
			(DmXk)		edge[bend right,swap,pos=0.0625]	node {$\Sigma\setminus\N{n}{k}$}	(t2);

\path[->]	(s1.west)	edge[out=0,in=90,looseness=1.2]	(s);
\path[->]	(s2.west)	edge[out=0,in=-90,looseness=1.2]	(s);

\begin{scope}[out=-90,in=180,looseness=0.4,very near start,swap]
\path	(D1X1)		edge	node {$\N{1}{1}$}	(s2.center)
		(D1Xk)		edge[looseness=1]	node {$\N{1}{k}$}	(s2.center);
\end{scope}
\begin{scope}[out=90,in=180,looseness=0.4,very near start]
\path	(DmX1)		edge	node {$\N{n}{1}$}	(s2.center)
		(DmXk)		edge[looseness=1]	node {$\N{n}{k}$}	(s2.center);
\end{scope}

\path[->]	(t2)	edge	node[swap,near end] {$\Sigma$}	(s);

\end{tikzpicture}
\caption{\label{fig:reduction-2}Reducing \SATUNSAT to \ShortestRW}
\end{figure}

Clearly, $\A$ can be constructed in polynomial time from $\phi$ and~$\psi$.
To establish our reduction, it remains to verify 1.--3.

To prove 1., assume that $\phi$ and~$\psi$ are both satisfiable. Since $\phi$
and~$\psi$ share no variable, there exists an assignment
$\alpha\colon\{X_1,\ldots,X_k\}\to\{\true,\false\}$ that satisfies both $\phi$
and~$\psi$. We claim that the word $01w$, where $w=w_1\ldots w_k\in\{0,1\}^k$
is defined by $w_j=1\Leftrightarrow \alpha(X_j)=\true$, resets~$\A$ to~$s$.
Clearly, $\delta^*(q,w)=s$ for all states~$q$ that are not of the form
$q=\P{i}{\bot}$, $q=\P{i}{\top}$, $q=\Q{i}{\bot}$ or $q=\Q{i}{\top}$.
Since $\delta^*(\P{i}{\bot},01)=\delta^*(\P{i}{\top},01)=\P{i}{1}$ and
$\delta^*(\Q{i}{\bot},01)=\delta^*(\Q{i}{\top},01)=\Q{i}{1}$ for each
$i=1,\ldots,n$, it suffices to show that $\delta^*(\P{i}{1},w)=
\delta^*(\Q{i}{1},w)=s$ for all~$i$. To prove that $\delta^*(\P{i}{1},w)=s$,
consider the least~$j$ such that either
$X_j\in C_i$ and $\alpha(X_j)=\true$ or $\neg X_j\in C_i$ and $\alpha(X_j)=
\false$ (such~$j$ exists since $\alpha$~satisfies~$\phi$). We have
$\delta^*(\P{i}{1},w_1\ldots w_{j-1})=\P{i}{j}$ and $\delta(\P{i}{j},w_j)
=s$ and therefore also $\delta^*(\P{i}{1},w)=s$.
The argument for $\delta^*(\Q{i}{1},w)=s$ is analogous.

Towards proving 2., assume that $\phi$~is satisfiable but $\psi$~is not.
Consider an assignment $\alpha\colon\{X_1,\ldots,X_k\}\to\{\true,\false\}$ that
satisfies~$\phi$. It follows with the same reasoning as above that
the word $01w1$, where $w\in\{0,1\}^k$ is defined by $w_j=1
\Leftrightarrow\alpha(X_j)=\true$, resets~$\A$ to~$s$.

To show that a \emph{shortest} reset word has length~$k+3$, it remains to show
that there exists no reset word of length~$k+2$. Towards a contradiction,
assume that $w=w_1\ldots w_{k+2}$~is such a word. Note that $w$~resets~$\A$
to~$s$ and that there exists $l\geq 2$ such that
$\delta^*(\Q{i}{\bot},w_1\ldots w_l)=\Q{i}{1}$ and
$\delta^*(\Q{i}{1},w_{l+1}\ldots w_{k+2})=s$ for all $i=1,\ldots,n$. Define
$\alpha\colon\{X_1,\ldots,X_k\}\to\{\true,\false\}$ by setting
$\alpha(X_j)=\true\Leftrightarrow w_{l+j}=1$. Since
$l\geq 2$ but $\delta^*(\Q{i}{1},w_{l+1}\ldots w_{k+2})=s$, for each~$i$
there must exist $j\in\{1,\ldots,k\}$ such that $\delta(\Q{i}{j},w_{l+j})=s$.
But then either $X_j\in D_i$ and $\alpha(X_j)=\true$ or $\neg X_j\in D_i$ and
$\alpha(X_j)=\false$. Hence, $\alpha$ is a satisfying assignment for~$\psi$,
contradicting our assumption that $\psi$ is unsatisfiable.

Finally, assume that $\phi$~is unsatisfiable. With the same reasoning as in
the previous case, it follows that there is no reset word
of length~$k+3$.
\qednow
\end{proof}

The above reduction shows \DP-hardness for an alphabet size of $|\Sigma|=2$.
For the special case of only one input letter, note that each reset word is of
the form $1^n$ for some~$n$. Asking whether there exists a reset word of
length~$k$ thus collapses to the question whether $1^k$ is a reset word
for~$\A$. This property can be decided with logarithmic space. Hence, both
problems, \ShortRW and \ShortestRW, are in \LOGSpace for $|\Sigma|=1$.

\section{Search problems}

In this section, we leave the realm of decision problems and enter the
(rougher) territory of search problems, where the task is not only to decide
whether a reset word of some length exists, but to compute a suitable word
(or its length).
More precisely, we deal with the following search problems:
\begin{itemize}
\item
Given a DFA~$\A$ and a positive integer~$k$ in unary,
compute a reset word for~$\A$ of length~$k$.
\item
Given a DFA~$\A$,
compute the length of a shortest reset word for~$\A$.
\item
Given a DFA~$\A$,
compute a shortest reset word for~$\A$.
\end{itemize}

Let us start with the first problem of computing a reset word of a given length.
It turns out that this problem is complete for the class \FNP of search problems
where the underlying binary relation is both polynomially balanced and
decidable in polynomial time.

\begin{proposition}\label{thm:fnp-complete}
The problem of computing a reset word of a given length is \FNP-complete.
\end{proposition}
\begin{proof}
Membership in \FNP follows from the fact that the binary relation
\[\{((\A,1^k),w)\mid\text{$w$ is a reset word for~$\A$ of length~$k$}\}\]
is polynomially balanced and polynomial-time decidable.

To prove hardness, we reduce from \FSAT, the problem of computing a satisfying
assignment for a given Boolean formula in conjunctive normal form. To this
end, we describe
two polynomial-time computable functions $f$ and~$g$, where $f$~computes
from a CNF formula~$\phi$ a synchronising automaton~$\A=f(\phi)$ over the
alphabet $\{0,1\}$ and a unary number~$k\in\bbN$, and $g$~computes from~$\phi$
and $w\in\Sigma^*$ an assignment for~$\phi$, such that, if $w$~is a reset word
for~$\A$ of length~$k$, then the generated assignment satisfies~$\phi$.

Eppstein~\cite{Eppstein90} showed how to compute in polynomial time, given a
CNF formula~$\phi=C_1\wedge\ldots\wedge C_n$ over the variables
$X_1,\ldots,X_k$, an automaton~$\A_\phi$ over the alphabet $\{0,1\}$ with the
following two properties:
\begin{enumerate}
	\item A word $w=w_1\cdots w_k$ is a reset word for~$\A$ if and only if the
assignment~$\alpha$, defined by $\alpha(X_j)=\true\Leftrightarrow w_j=1$,
satisfies~$\phi$.
	\item An assignment $\alpha\colon\{X_1,\ldots,X_k\}\to\{\true,\false\}$
satisfies $\phi$ if and only if the word $w\in\{0,1\}^k$, defined by
$w_j=1 \Leftrightarrow \alpha(X_j)=\true$, is a reset word for~$\A$.
\end{enumerate}
(Note that the reduction we use to prove \cref{thm:dp-complete} has similar
properties and could also be used.)

Hence, we can choose $f$ to be the function that maps~$\phi$ to~$(\A_\phi,1^k)$
and $g$~to be the function that maps $(\phi, w)$ to the corresponding assignment
$\alpha$. (If $|w|\neq k$, then $\alpha$~can be chosen arbitrarily.)
\qednow
\end{proof}

\begin{remark}
Note that the mapping $f:\{0,1\}^k\to\{\true,\false\}^{\{X_1,\ldots,X_k\}}$,
defined by $f(w)(X_j)=\true\Leftrightarrow w_j=1$, is a bijection. Eppstein's
reduction shows that one can compute from a Boolean formula~$\phi$ over the
variables $\{X_1,\dots,X_k\}$ an automaton~$\A$ such that $f$~remains a
bijection when one restricts the domain to reset words for~$\A$ and the range
to assignments that satisfy~$\phi$. Therefore, his reduction can be viewed as a
parsimonious reduction from \SharpSAT, the problem of counting all satisfying
assignments of a given Boolean formula, to the problem of counting all reset
words of a given length (represented in unary).
Since the first problem is complete for \SharpP \cite{Valiant79},
the second problem is \SharpP-hard. On the other hand, it is easy to see
that the second problem is in \SharpP. Hence, this problem is
\SharpP-complete.
\end{remark}

Next, we consider the problem of computing the \emph{length} of a shortest reset
word for a given automaton: we establish that this problem is complete for the
class \FPNPlog of all problems that are solvable by a polynomial-time algorithm
with access to an oracle for a problem in \NP where the number of queries is
restricted to $\Oh(\log n)$.

\begin{theorem}\label{thm:fpnplog-complete}
The problem of computing the length of a shortest reset word is
\FPNPlog-complete.
\end{theorem}
\begin{proof}
To prove membership in \FPNPlog, consider \cref{alg:1} which is a
bi\-na\-ry-search
algorithm for determining the length of a shortest reset word for an
automaton~$\A$ with $n$~states.
The algorithm is executed in polynomial time: the while~loop is repeated
$\Oh(\log n)$ times and asks $\Oh(\log n)$ queries to the oracle, which is
used for determining whether $\A$~has a reset word of a given length.

\begin{algorithm}
\vspace*{.8ex}
\begin{tabbing}
\hspace*{1em}\=\hspace{1em}\=\hspace{1em}\=\hspace{1em}\=
\hspace{1em}\=\hspace{1em}\= \kill

\textbf{if} $\A$~is not synchronising \textbf{then reject} \\
$\low\coloneq -1$ \\
$\high\coloneq(n^3-n)/6$ \\
\+\textbf{while} $\high-\low>1$ \textbf{do} \\
$k\coloneq\lceil(\low+\high)/2\rceil$ \\
\+\textbf{if} $\A$ has a reset word of length~$k$ \textbf{then} \\
\-$\high\coloneq k$ \\
\+\textbf{else} \\
\-\-$\low\coloneq k$ \\
\textbf{end while} \\
\textbf{return} $\high$
\end{tabbing}
\vspace*{-2ex}
\caption{\label{alg:1}Computing the length of a shortest reset word}
\end{algorithm}

Krentel \cite{Krentel88} showed that \MAXSATSIZE, the problem of computing
the maximum number of simultaneously satisfiable clauses of a CNF formula, is
complete for \FPNPlog. Therefore, to
establish \FPNPlog-hardness, it suffices to give a reduction from
\MAXSATSIZE to our problem. Such a reduction consists of
two polynomial-time computable functions $f$ and~$g$ with the following
properties: $f$~computes from a CNF formula~$\phi$ a (synchronising)
automaton~$\A=f(\phi)$, and $g$~computes
from~$\phi$ and $l\in\bbN$ a new number $g(\phi,l)\in\bbN$ such that, if
$l$~is the length of a shortest reset word for~$\A$, then the maximum number of
simultaneously satisfiable clauses in~$\phi$ equals~$g(\phi,k)$.

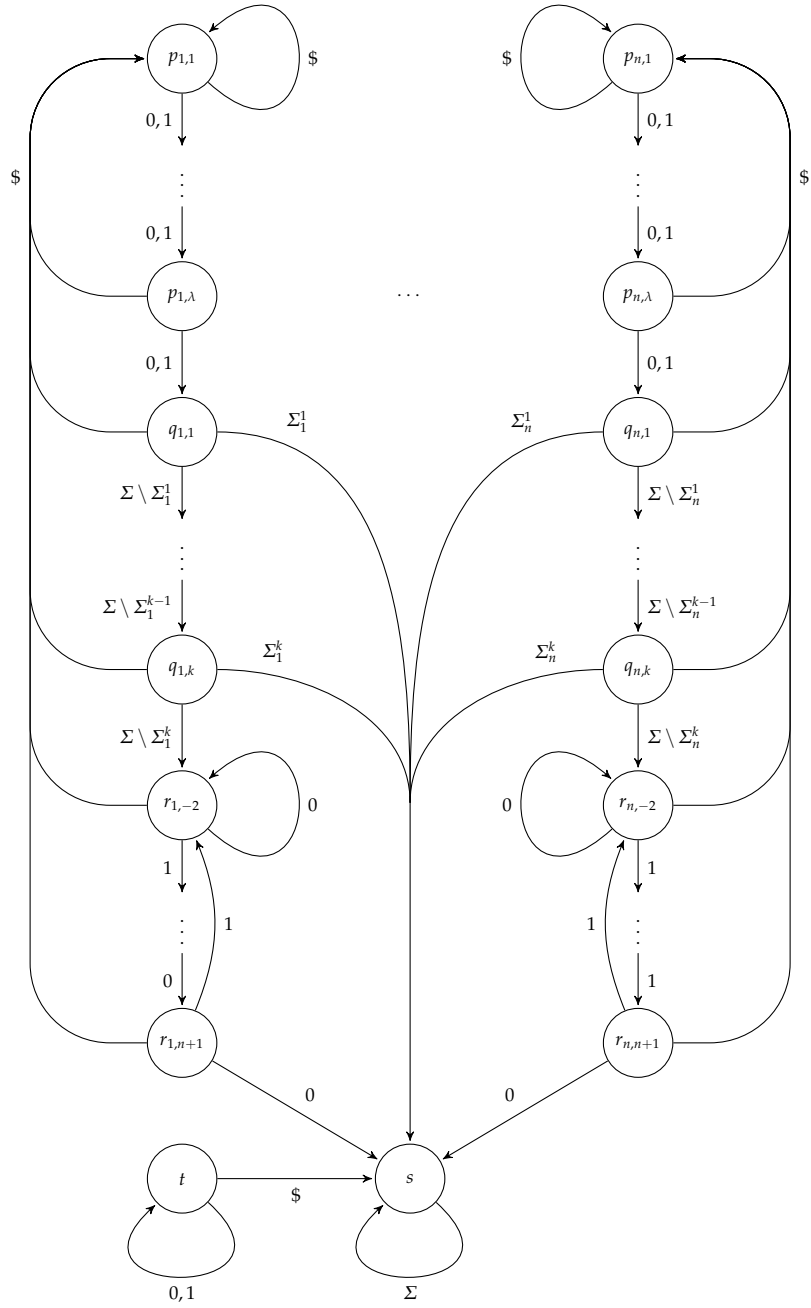
\begin{figure}
\centering
\begin{tikzpicture}[yscale=0.9,automaton]

\node[state]	(C1p1) at (0,5.5)		{\P{1}{1}};
\node			(C1p2) at (0,3.75)		{\vdots};
\node[state]	(C1pl) at (0,2)			{\P{1}{\lambda}};
\node[state]	(C1X1) at (0,0)			{\Q{1}{1}};
\node			(C1X2) at (0,-1.75)		{\vdots};
\node[state]	(C1Xk) at (0,-3.5)		{\Q{1}{k}};
\node[state]	(C1r-2) at (0,-5.5)	{\R{1}{-2}};
\node			(C1r2) at (0,-7.25)		{\vdots};
\node[state]	(C1rBOT) at (0,-9)	{\R{1}{n+1}};

\node at (3,2) {$\cdots$};

\begin{scope}[xshift=6cm]
\node[state]	(Cnp1)	at (0,5.5)		{\P{n}{1}};
\node			(Cnp2)	at (0,3.75)		{\vdots};
\node[state]	(Cnpl)	at (0,2)		{\P{n}{\lambda}};
\node[state]	(CnX1)	at (0,0)		{\Q{n}{1}};
\node			(CnX2)	at (0,-1.75)	{\vdots};
\node[state]	(CnXk)	at (0,-3.5)		{\Q{n}{k}};
\node[state]	(Cnr-2) at (0,-5.5)		{\R{n}{-2}};
\node			(Cnr2)	at (0,-7.25)	{\vdots};
\node[state]	(CnrBOT)	at (0,-9)	{\R{n}{n+1}};
\end{scope}

\node			(s1) at (3,-5.5)	{};

\node[state]	(t) at (0,-11)		{$t$};
\node[state]	(s) at (3,-11)		{$s$};


\path[->,swap]	(C1p1)		edge	node {$0,1$}						(C1p2)
			(C1p2)		edge	node {$0,1$}						(C1pl)
			(C1pl)		edge	node {$0,1$}						(C1X1)
			(C1X1)		edge	node {$\Sigma\setminus\M{1}{1}$}	(C1X2)
			(C1X2)		edge	node {$\Sigma\setminus\M{1}{k-1}$}	(C1Xk)
			(C1Xk)		edge	node {$\Sigma\setminus\M{1}{k}$}	(C1r-2)
			(C1r-2)		edge	node {$1$}							(C1r2)
			(C1r2)		edge	node {$0$}							(C1rBOT)
			(C1rBOT)	edge[bend right]	node {$1$}				(C1r-2);

\path[->]	(Cnp1)		edge	node {$0,1$}						(Cnp2)
			(Cnp2)		edge	node {$0,1$}						(Cnpl)
			(Cnpl)		edge	node {$0,1$}						(CnX1)
			(CnX1)		edge	node {$\Sigma\setminus\M{n}{1}$}	(CnX2)
			(CnX2)		edge	node {$\Sigma\setminus\M{n}{k-1}$}	(CnXk)
			(CnXk)		edge	node {$\Sigma\setminus\M{n}{k}$}	(Cnr-2)
			(Cnr-2)		edge	node {$1$}							(Cnr2)
			(Cnr2)		edge	node {$1$}							(CnrBOT)
			(CnrBOT)	edge[bend left]	node {$1$}					(Cnr-2);

\begin{scope}[rounded corners=3em,->]
\draw	(C1rBOT) -- +(-2,0) -- (-2,5.5) -- (C1p1);
\draw	(C1r-2) -- +(-2,0) -- (-2,5.5) -- (C1p1);
\draw	(C1Xk) -- +(-2,0) -- (-2,5.5) -- (C1p1);
\draw	(C1X1) -- +(-2,0) -- (-2,5.5) -- (C1p1);
\draw	(C1pl) -- +(-2,0) -- node {$\$$} (-2,5.5) -- (C1p1);
\end{scope}
\path[->] (C1p1) edge[out=-45,in=45,loop] node[swap] {$\$$} (C1p1);

\begin{scope}[rounded corners=3em,->]
\draw	(CnrBOT) -- +(2,0) -- (8,5.5) -- (Cnp1);
\draw	(Cnr-2) -- +(2,0) -- (8,5.5) -- (Cnp1);
\draw	(CnXk) -- +(2,0) -- (8,5.5) -- (Cnp1);
\draw	(CnX1) -- +(2,0) -- (8,5.5) -- (Cnp1);
\draw	(Cnpl) -- +(2,0) -- node[swap] {$\$$} (8,5.5) -- (Cnp1);
\end{scope}
\path[->] (Cnp1) edge[out=-135,in=135,loop] node {$\$$} (Cnp1);


\begin{scope}[out=0,in=90,looseness=1,very near start]
\path	(C1X1)		edge	node {$\M{1}{1}$}	(s1.center)
		(C1Xk)		edge	node {$\M{1}{k}$}	(s1.center);
\end{scope}
\begin{scope}[out=180,in=90,looseness=1,very near start,swap]
\path	(CnX1)		edge	node {$\M{n}{1}$}	(s1.center)
		(CnXk)		edge	node {$\M{n}{k}$}	(s1.center);
\end{scope}
\path[->]	(s1.north)		edge	(s);

\path[->]	(C1r-2)	edge[out=-45,in=45,loop]	node[swap] {$0$}		(C1r-2)
			(Cnr-2)	edge[out=-135,in=135,loop]	node {$0$}		(Cnr-2);
\path[->]	(C1rBOT)	edge	node {$0$}					(s)
			(CnrBOT)	edge	node[swap] {$0$}							(s);
\path[->]	(t)	edge[out=-45,in=-135,loop]		node {$0,1$}			(t)
			(t)	edge							node[swap] {$\$$}		(s);
\path[->]	(s)	edge[out=-45,in=-135,loop]		node {$\Sigma$}			(s);

\end{tikzpicture}
\caption{\label{fig:reduction-3}Reducing \MAXSATSIZE to computing the length of
a shortest reset word}
\end{figure}

Given a formula~$\phi=C_1\wedge\ldots\wedge C_n$ over propositional
variables $X_1,\ldots,X_k$, the resulting automaton~$\A$ is depicted in
\cref{fig:reduction-3}:
The input alphabet is $\Sigma\coloneq\{0,1,\$\}$,
and the sets $\M{i}{j}\subseteq\Sigma$ are defined as in the proof of
\cref{thm:dp-complete}; we set $\lambda\coloneq k+n(n+4)$.
The behaviour of the transition function on vertices of the form~$\R{i}{j}$
is defined as follows:
\begin{itemize}
  \item $\delta(\R{i}{j},\$)=\P{i}{1}$ for all $j\in\{-2,\dots,n+1\}$;
  \item $\delta(\R{i}{j},1)=\R{i}{j+1}$, $\delta(\R{i}{j},0)=\R{i}{-2}$ for
all $j\in\{-2,-1,i\}$;
  \item $\delta(\R{i}{j},1)=\R{i}{-2}$, $\delta(\R{i}{j},0)=\R{i}{j+1}$ for
all $j\in\{0,\dots,i-1,i+1,\dots,n\}$;
  \item $\delta(\R{i}{n+1},1)=\R{i}{-2}$, $\delta(\R{i}{n+1},0)=s$.
\end{itemize}


It is not difficult to see that $\A$~can be constructed in polynomial time
from~$\phi$. Moreover, we claim that, for each $m\in\{0,1,\ldots,n\}$, there
exists an assignment that satisfies
at least $n-m$~clauses of~$\phi$ if and only if $\A$~has a reset word of length
$1+\lambda+k+m(n+4)$. Hence, if $l$~is the length of a shortest reset word
for~$\A$, then the maximal number of simultaneously satisfiable clauses of~$\phi$
is given by $n-\left\lceil\frac{\max\{0,l-1-\lambda-k\}}{n+4}\right\rceil$.
Clearly, this number can be computed in polynomial time from~$\phi$ and~$l$.

($\Rightarrow$) Assume that $\alpha\colon\{X_1,\ldots,X_k\}\to\{\true,\false\}$
is an assignment that satisfies all clauses except (possibly) the clauses
$C_{i_1},\ldots,C_{i_m}$, and consider the word
\[w\coloneq\$ 1^\lambda x_1\ldots x_k z_{i_1}\ldots z_{i_m},\]
where $z_i=110^i 10^{n-i+1}\in\{0,1\}^{n+4}$ for
$i\in\{1,\ldots,n\}$ and
\[x_j\coloneq\begin{cases}
    1 & \text{if $\alpha(X_j)=\true$}, \\
    0 & \text{otherwise.}
  \end{cases}\]
Note that $w$~has length $1+\lambda+k+m(n+4)$. We claim that
$w$~resets~$\A$ to~$s$. Since reading~$\$$ has the effect of going from each
state of the form $\P{i}{j}$, $\Q{i}{j}$ or $\R{i}{j}$ to $\P{i}{1}$ and from
$t$ to~$s$, and reading $1^\lambda$ has the effect of going
from $\P{i}{1}$ to $\Q{i}{1}$, it suffices to show that
$\delta^*(\Q{i}{1},x_1\ldots x_k z_{i_1}\ldots z_{i_m})=s$. If $C_i$~is
satisfied by~$\alpha$, then this follows from the fact that there exists~$j$
such that $\delta(\Q{i}{j},x_j)=s$. Otherwise, we have
$\delta^*(\Q{i}{1},x_1\ldots x_k)=\R{i}{-2}$, $\delta^*(\R{i}{-2},z_j)
=\R{i}{-2}$
for all $j\not=i$, but $\delta^*(\R{i}{-2},z_i)=s$. Since
$i\in\{i_1,\ldots,i_m\}$, this implies that $\delta^*(\Q{i}{1},x_1\ldots x_k
z_{i_1}\ldots z_{i_m})=s$.

($\Leftarrow$) Assume that $\A$~has a reset word of length
$1+\lambda+k+m(n+4)$, and let $w$ be a shortest reset word for~$\A$.
We claim that $w$~has the form $w=\$u$ or $w=u\$$ for $u\in\{0,1\}^*$.
Otherwise, $w=u\$v$ for $u,v\in\Sigma^+$. Towards a contradiction, we
distinguish the following two cases: $|u|\leq\lambda$ and $|u|>\lambda$.
If $|u|\leq\lambda$, then $\delta^*(\P{i}{1},u\$)=\P{i}{1}$ for all
$i=1,\ldots,n$, and the word $\$v$ would be a shorter reset word than~$w$.
Now assume that $|u|>\lambda$. It must be the case that
$\delta^*(\P{i}{1},u)\neq s$ for some $i\in\{1,\ldots,n\}$ because otherwise
$\$u$~would be a shorter reset word than~$w$. But then
$\delta^*(\P{i}{1},u\$)=\P{i}{1}$. Hence, since $w$~resets~$\A$
to~$s$ and the shortest path from $\P{i}{1}$ to~$s$ has length greater
than~$\lambda$, $|v|>\lambda$ and $|w|>1+2\lambda\geq 1+\lambda+k+n(n+4)
\geq 1+\lambda+k+m(n+4)$, a contradiction.

Now, if $\phi$ is satisfiable, we are done. Otherwise, let us fix
$u\in\{0,1\}^*$ such that $w=\$u$ or $w=u\$$. Since $\phi$ is not
satisfiable, $|u|\geq\lambda+k$.
Let $u=y x_1\ldots x_k z$ where $y,z\in\{0,1\}^*$, $|y|=\lambda$,
and $x_j\in\{0,1\}$ for all $j=1,\ldots,k$. Now consider the
assignment~$\alpha$ defined by
\[\alpha(X_j)=\begin{cases}
  \true & \text{if $x_j=1$,} \\
  \false & \text{otherwise.}
\end{cases}\]
Moreover, let
\[I\coloneq\{i\in\{1,\ldots,n\}\ |\ 
  \text{$C_i$ is not satisfied by~$\alpha$}\}.\]
We claim that $|I|\leq m$ (so $\alpha$~satisfies at least $n-m$ clauses
of~$\phi$). To see this, first note that $\delta^*(\P{i}{1},yx_1\ldots x_k)=
\R{i}{-2}$ for all $i\in I$. Hence, we must have that
$\delta^*(\R{i}{-2},z)=s$ for all such~$i$. By the construction of~$\A$,
this is only possible if $z$~contains the word $110^i10^{n-i+1}$ as an infix
for each $i\in I$.
Since these infixes cannot overlap, $|z|\geq |I|\cdot (n+4)$. On the other
hand, since $|u|\leq\lambda+k+m(n+4)$, we must have $|z|\leq m(n+4)$.
Hence, $|I|\leq m$.
\qednow
\end{proof}

The construction we have presented to prove \cref{thm:fpnplog-complete} uses a
three-letter alphabet. With a little more effort, we can actually reduce
the alphabet to an alphabet with two letters $0$ and~$1$: For each
state~$q\notin\{s,t\}$ of~$\A$, there are three
states $(q,0)$, $(q,1)$ and $(q,2)$ in the new automaton~$\A'$. Additionally,
$\A'$~contains the states $(t,0)$, $(t,1)$ and~$s$. The new transition
function~$\delta'$ is defined as follows:
\begin{align*}
  & \delta'((q,0),0)=(q,1), && \delta'((q,0),1)=(q,2), \\ 
  & \delta'((q,1),0)=(q,1), && \delta'((q,1),1)=(\delta(q,\$),2), \\
  & \delta'((q,2),0)=(\delta(q,0),0), && \delta'((q,2),1)=(\delta(q,1),0)
\intertext{for all $q\notin\{s,t\}$, and}
  & \delta'((t,0),0)=s, && \delta'((t,0),1)=(t,1), \\
  & \delta'((t,1),0)=(t,0), && \delta'((t,1),1)=(t,1), \\
  & \delta'(s,0)= s, && \delta'(s,1)=s.
\end{align*}
Intuitively, taking a transition in~$\A$ corresponds to taking two
transitions in~$\A'$. It is not difficult to see that a shortest reset
word for~$\A'$ has length~$2l$ if a shortest reset word for~$\A$ has length~$l$.

For the potentially harder problem of computing a shortest reset word (not only
its length), we can only prove membership in \FPNP, the class of all search
problems that are solvable in polynomial time using an oracle for a problem
in \NP (without
any restriction on the number of queries). Of course, hardness for
\FPNPlog carries over from our previous result. We have not been able to close
the gap between the two bounds. To the best of our knowledge, the same situation
occurs e.g.\ for \MAXSAT, where the aim is to find an assignment of a given
Boolean formula that satisfies as many clauses as possible.

\begin{theorem}
The problem of computing a shortest reset word is in \FPNP and hard for
\FPNPlog.
\end{theorem}
\begin{proof}
To prove membership in \FPNP, consider \cref{alg:2} for computing a
shortest reset word for an automaton~$\A$ over any finite
alphabet~$\Sigma$.
The algorithm obviously computes a reset word of length~$l$, which is the length of
a shortest reset word. To see that the algorithm runs in polynomial time if it has
access to an \NP oracle, note that deciding whether $\A$~has a reset word of a
given length with a given prefix is in \NP (since a nondeterministic
polynomial-time algorithm can guess such a word). Moreover, as we have shown
above, computing the \emph{length} of a shortest reset word can be done by a
polynomial-time algorithm with access to an \NP oracle.

\begin{algorithm}
\vspace*{.8ex}
\begin{tabbing}
\hspace*{1em}\=\hspace{1em}\=\hspace{1em}\=\hspace{1em}\=
\hspace{1em}\=\hspace{1em}\= \kill

\textbf{if} $\A$ is not synchronising \textbf{then reject} \\
Compute the length~$l$ of a shortest reset word for~$\A$ \\
$w\coloneq\epsilon$ \\
\+\textbf{while} $|w|<l$ \textbf{do} \\
\+\textbf{for each} $a\in\Sigma$ \textbf{do} \\
\+\textbf{if} $\A$ has a reset word of length~$l$ with prefix~$wa$
  \textbf{then} \\
\-$w\coloneq wa$; \textbf{break for} \\
\-\textbf{end if} \\
\-\textbf{end for} \\
\textbf{end while} \\
\textbf{return} $w$
\end{tabbing}
\vspace*{-2ex}
\caption{\label{alg:2}Computing a shortest reset word}
\end{algorithm}

Hardness for \FPNPlog follows from \cref{thm:fpnplog-complete} since the
problem of computing the length of a shortest reset word is trivially
reducible to the problem of computing a shortest reset word: an instance of
the former problem is also an instance of the latter problem, and a
solution of the latter problem can be turned into a solution of the former
problem by computing its length.
\qednow
\end{proof}

\section{Conclusion}

\begin{figure}[t]
\begin{tikzpicture}[x=1.45cm,y=0.6cm,label distance=-2pt]
	\begin{scope}
	\useasboundingbox (0,-3.2) rectangle (4,10);
	\node at (2,1.2) {\PTime};
	\node at (0.95,2.5) {\NP};
	\node at (2.9,2.7) {\coNP};
	\node at (0.7,4.3) {\DP};
	\node at (3.35,4.3) {\coDP};
	\node at (2,7.4) {\PNPlog};
	\node at (2,9) {\PNP};

	\draw (0,0) -- (4,0); 
	\draw (0,0) .. controls (2,3) .. (4,0); 
	\draw (0,0) .. controls (1.2,5) .. (4,0); 
	\draw (0,0) .. controls (2.8,5) .. (4,0); 
	\draw (0,0) .. controls (0.2,10.5) and (2.8,7) .. (4,0); 
	\draw (0,0) .. controls (1.2,7) and (3.8,10.5) .. (4,0); 
	\draw (0,0) .. controls (0.1,11) and (3.9,11) .. (4,0); 
	\draw (0,0) .. controls (0.1,13) and (3.9,13) .. (4,0); 

	\node (b1) [bullet] at (1.4,3) {};
	\node (b2) [bullet] at (1.2,5.8) {};
	
	\node (p1) [anchor=south west,text depth=0pt] at (1.5,-1) {\ShortRW};
	\node (p2) [anchor=south west,text depth=0pt] at (1.5,-1.8) {\ShortestRW};
	
	\draw (b1) |- (p1); \draw (b2) |- (p2);
	\end{scope}

	\begin{scope}[xshift=6.3cm]
	\useasboundingbox (0,-3.2) rectangle (4,10);
	\node at (2,1.2) {\FP};
	\node at (1.3,6.5) {\FNP};
	\node at (2.65,6.9) {\FPNPlog};
	\node at (2,9) {\FPNP};

	\draw (0,0) -- (4,0); 
	\draw (0,0) .. controls (2,3) .. (4,0); 
	\draw (0,0) .. controls (1,10) and (1,10) .. (4,0); 
	\draw (0,0) .. controls (3,11) and (3,11) .. (4,0); 
	\draw (0,0) .. controls (0.1,13) and (3.9,13) .. (4,0); 

	\node (b1) [bullet] at (0.9,6.0) {};
	\node (b2) [bullet] at (3.1,6.0) {};
	\node (b3) [bullet] at (3.3,6.7) {};
	
	\node (p1) [anchor=south east,text depth=0pt] at (3,-1) {compute short reset word};
	\node (p2a) [anchor=south east,text depth=0pt] at (3,-1.8) {compute length of};
	\coordinate (p2ae) at (p2a.east);
	\node (p2b) [anchor=south east,text depth=0pt] at (3,-2.4) {shortest reset word};
	\coordinate (p2be) at (p2b.east);
	\coordinate (p2) at (barycentric cs:p2ae=1,p2be=1);
	\node (p3) [anchor=south east,text depth=0pt] at (3,-3.2) {compute shortest reset word};
	
	\draw (b1) -- (b1 |- p1.north);
	\draw (b2) |- (p2); \draw (b3) |- (p3);
	\end{scope}
\end{tikzpicture}
\caption{\label{fig:summary}Summary of results}
\end{figure}
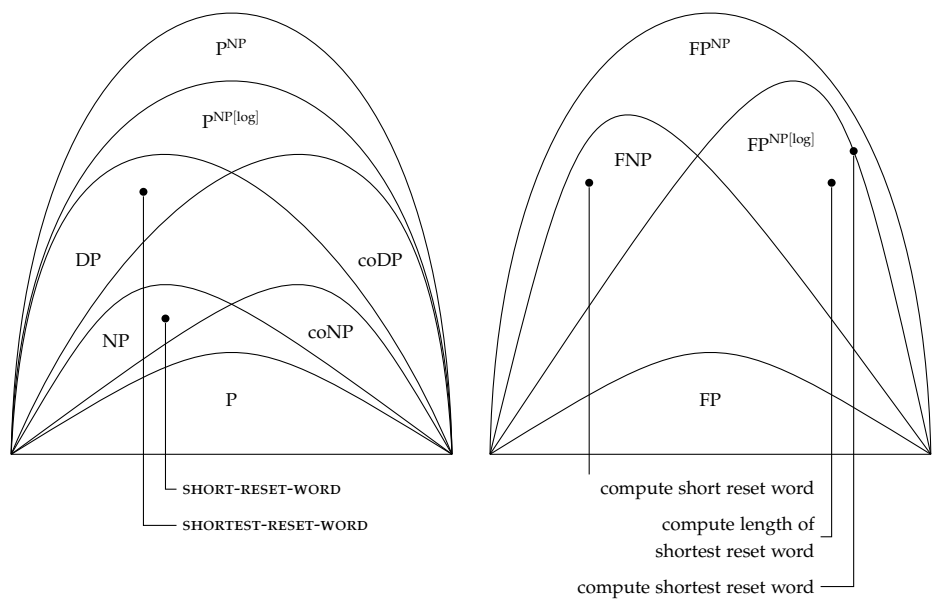

We have investigated several decision problems and search problems about finding
reset words in finite automata.
The results we have obtained shed more light on the difficulty of computing
such words. In particular, deciding whether for a given automaton a shortest
reset word has length~$k$ is \DP-complete, and computing the length of a
shortest reset word is \FPNPlog-complete, i.e.\ as hard as calculating the
maximum number of simultaneously satisfiable clauses of a Boolean formula.
A summary of all our results is depicted in \cref{fig:summary}. (See
\cite{Papadimitriou94,Selman94} for the relationships between the referred
complexity classes.)

\subsubsection{Acknowledgements}

We thank an anonymous reviewer for pointing out \cite{Gawrychowski08}. Moreover,
we are grateful to Christof L\"oding and Wolfgang Thomas for helpful comments on
an early draft of this paper.

\bibliographystyle{abbrv}
\bibliography{cerny.bib}

\begin{thebibliography}{10}

\bibitem{Berlinkov10}
M.~V. Berlinkov.
\newblock Approximating the length of synchronizing words.
\newblock In {\em Proceedings of the 5th International Computer Science
  Symposium in Russia, CSR~2010}, volume 6072 of {\em Lecture Notes in Computer
  Science}, pages 37--47. Springer-Verlag, 2010.

\bibitem{Cerny64}
J.~{\v{C}}ern\'{y}.
\newblock Pozn{\'a}mka k. homog{\'e}nnym experimentom s konecn{\'y}mi
  automatmi.
\newblock {\em Matematicko-fyzikalny {\v C}asopis Slovensk.\ Akad.\ Vied},
  14(3):208--216, 1964.

\bibitem{CPR71}
J.~{\v{C}}ern\'{y}, A.~Pirick\'{a}, and B.~Rosenauerov\'{a}.
\newblock On directable automata.
\newblock {\em Kybernetica}, 7(4):289--298, 1971.

\bibitem{Eppstein90}
D.~Eppstein.
\newblock Reset sequences for monotonic automata.
\newblock {\em {SIAM} Journal on Computing}, 19(3):500--510, 1990.

\bibitem{Gawrychowski08}
P.~Gawrychowski.
\newblock Complexity of shortest synchronizing word.
\newblock Unpublished man\-u\-script, April 2008.

\bibitem{Krentel88}
M.~W. Krentel.
\newblock The complexity of optimization problems.
\newblock {\em Journal of Computer and System Sciences}, 36:490--509, 1988.

\bibitem{Papadimitriou94}
C.~H. Papadimitriou.
\newblock {\em Computational complexity}.
\newblock Addison-Wesley, 1994.

\bibitem{Pin83}
J.-{\'E}. Pin.
\newblock On two combinatorial problems arising from automata theory.
\newblock {\em Annals of Discrete Mathematics}, 17:535--548, 1983.

\bibitem{Samotij07}
W.~Samotij.
\newblock A note on the complexity of the problem of finding shortest
  synchronizing words.
\newblock In {\em Proceedings of AutoMathA 2007}. University of Palermo (CD),
  2007.

\bibitem{Selman94}
A.~L. Selman.
\newblock A taxonomy of complexity classes of functions.
\newblock {\em Journal of Computer and System Sciences}, 48(2):357--381, 1994.

\bibitem{Valiant79}
L.~G. Valiant.
\newblock The complexity of computing the permanent.
\newblock {\em Theoretical Computer Science}, 8:189--201, 1979.

\bibitem{Volkov08}
M.~V. Volkov.
\newblock Synchronizing automata and the \v{C}ern\'{y} conjecture.
\newblock In {\em Proceedings of the 2nd International Conference on Language
  and Automata Theory and Applications, LATA~2008}, volume 5196 of {\em Lecture
  Notes in Computer Science}, pages 11--27. Springer-Verlag, 2008.

\end{thebibliography}

\iffull

\clearpage
\appendix
\section{Complexity classes}
\label{appendix:complexity}

In this appendix, we want to give an overview on the complexity classes that
play a role in this paper. 
More detailed information can be found in the literature
\cite{Papadimitriou94,Selman94}.

\subsection{Decision problems}

We assume that the reader is familiar with the complexity classes
\PTime, \NP and \coNP.
The class \DP is the closure of $\NP\cup\coNP$ under intersection.
Equivalently, a language~$L$ is in \DP if and only if it is of the form
$L=L_1\setminus L_2$ with $L_1,L_2\in\NP$. The canonical complete
problem for \DP is the following problem, which is derived from \SAT.

\begin{quote}
\SATUNSAT: Given two Boolean formulae $\phi$ and~$\psi$ in CNF, decide whether
$\phi$~is satisfiable and $\psi$~is unsatisfiable.
\end{quote}

Note that, unless $\NP=\coNP$, \SATUNSAT is not an element of either \NP or
\coNP. Hence, it is conjectured that \DP is a proper superclass of
$\NP\cup\coNP$. On the other hand, any problem in \DP can be solved by
a polynomial-time algorithm that has access to an oracle for an \NP-complete
problem (for instance, \SAT). Hence, \DP is contained in $\PTime^\NP$, the first
level of the polynomial hierarchy.

\subsection{Counting problems}

Formally, a \emph{counting problem} is just a function $F\colon\Sigma^*\to\bbN$.
The class \SharpP consists of all counting problems for which there exists a
nondeterministic poly\-no\-mi\-al-time Turing machine $M$ such that for each
input~$x$
the number of accepting runs of~$M$ on~$x$ equals~$F(x)$.
Analogously to the verifier definition of \NP, \SharpP can also be characterised
in terms of a relation: we have $F\in\SharpP$ if and only if there
exists a polynomial-time decidable, polynomially balanced relation
$R\subseteq\Sigma^*\times\Sigma^*$ such that
$F(x)=|\{y\in\Sigma^*\mid(x,y)\in R\}|$. (A binary relation~$R$ is polynomially
balanced if there exists a polynomial~$p$ such that $|y|\leq p(|x|)$ for all
$(x,y)\in R$.)

The simplest (but also most restrictive) kind of a reduction between counting
problems is the parsimonious reduction. Formally, a
\emph{parsimonious reduction} from a counting problem $F\colon\Sigma^*\to\bbN$ to
another counting problem $G\colon\Sigma^*\to\bbN$ is a polynomial-time computable
function $f\colon\Sigma^*\to\Sigma^*$ such that $|F(x)|=|G(f(x))|$ for all
$x\in\Sigma^*$.
A function problem~$F$ is \SharpP-complete if $F\in\SharpP$ and for every
$G\in\SharpP$ there exists a parsimonious reduction from $G$ to~$F$.
(In the literature, \SharpP-hardness is often defined via polynomial-time
Turing reductions, which are more general than parsimonious reductions.)

The canonical \SharpP-complete problem is \SharpSAT where the number of
satisfying assignments for a given Boolean formula is sought. In fact,
\SharpSAT is \SharpP-hard even for formulae in conjunctive normal form with
only two literals per clause. (Note that \TwoSAT, the restriction of \SAT
to such formulae, is in \PTime)

\subsection{Search problems}

A more general concept for computational problems is the one of a search problem.
Formally, a \emph{search problem} is a binary
relation $R\subseteq\Sigma^*\times\Sigma^*$.
If $(x,y)\in R$, we say that $y$~is a \emph{solution} for~$x$ (wrt.~$R$).
An algorithm~$A$ \emph{solves} the search problem~$R$, if the following two
conditions hold for every input~$x\in\Sigma^*$:
\begin{itemize}
\item If $x$~has a solution, then $A$~accepts~$x$ and
outputs such a solution, i.e.\ a word $y\in\Sigma^*$ with $(x,y)\in R$.
\item If $x$~has no solution, then the algorithm rejects~$x$.
\end{itemize}

Any decision class that is defined with respect to a deterministic machine
model can be extended to a class of search problems in a straightforward way.
For example, the class \FP consists of all search problems solvable in
polynomial time by a deterministic Turing machine with a dedicated output tape.
By equipping the machine with an oracle for an \NP-complete problem, we obtain
the complexity classes \FPNP and \FPNPlog, depending on the allowed number of
queries. For \FPNP, the number of queries is not restricted, whereas for \FPNPlog
only a logarithmic number of queries (in the length of the input) is allowed.
Obviously, we have $\FP\subseteq\FPNPlog\subseteq\FPNP$.

For the class \NP, it turns out that there are two different classes
of search problems that can be derived from it. If one takes the usual definition
of \NP that refers to nondeterministic Turing machines, one arrives at the class
\NPMV of search problems that can be solved by a nondeterministic machine in
polynomial time (see \cite{Selman94} for a formal definition). On the other hand,
if one takes the ``verifier definition'' of \NP, one arrives at the class \FNP
(called \NPMVg in \cite{Selman94})
of search problems where the underlying binary relation~$R$ is both polynomially
balanced and decidable in polynomial time.

Although both classes \FNP and \NPMV are derived from the same class \NP, they
are not necessarily equal. It is easy to see that $\FNP\subseteq\NPMV$, but we
have $\NPMV\subseteq\FNP$ if and only if $\FP=\FNP$, which in turn is equivalent
to $\PTime=\NP$ \cite{Selman94}. Regarding other inclusions, both \FNP and \NPMV
are contained in \FPNP, but they are incomparable with \FPNPlog under the
assumption that $\NP\neq\coNP$ \cite{Selman94}.

As for decision and counting problems, one can compare search problems using
reductions. However, a reduction between two search problems does not only
consist of one function that maps instances to instances but also of another
function that maps solutions to solutions (or even a collection of such
functions, one for each instance). Formally, a reduction
from a search problem $R\subseteq\Sigma^*\times\Sigma^*$ to another search
problem $S\subseteq\Sigma^*\times\Sigma^*$ consists of two
polynomial-time computable functions $f\colon\Sigma^*\to\Sigma^*$ and
$g\colon\Sigma^*\times\Sigma^*\to\Sigma^*$ such that the
following two conditions hold for every $x,y\in\Sigma^*$:
\begin{itemize}
  \item If $x$~has a solution wrt.~$R$, then $f(x)$~has a solution wrt.~$S$.
  \item If $y$ is a solution for~$f(x)$ wrt.~$S$, then $g(x,y)$ is a solution
for~$x$ wrt.~$R$.
\end{itemize}

The canonical complete problem for \FNP is \FSAT, the
problem of computing a satisfying assignment for a given Boolean formula.
In fact, \FSAT is not only complete for \FNP, but also for \NPMV. This
does not imply $\PTime=\NP$ since \FNP is \emph{not} closed under
reductions unless $\PTime=\NP$:
Consider, for instance, the variant $\FSAT_1$ of \FSAT where one has to
output~$1$ if the formula is satisfiable instead of outputting a satisfying
assignment, i.e.\ $\FSAT_1\coloneq\{(\phi,1)\mid\phi\in\SAT\}$.
Clearly, there exists a reduction from $\FSAT_1$ to \FSAT, but $\FSAT_1$
is not contained in \FNP unless $\SAT\in\PTime$.

A problem that is complete for \FPNPlog is \MAXSATSIZE, where the task is
to compute the maximal number of clauses that are satisfiable simultaneously
for a given Boolean formula in conjunctive normal form. Finally, if clauses
are equipped with weights and an assignment that maximises the total weight
of satisfied clauses is sought, one arrives at \MAXWEIGHTSAT, a problem that
is complete for \FPNP \cite{Krentel88}.

\fi

\end{document}